\documentclass[apjl]{emulateapj}

\shorttitle{The Nature of LMC-5}
\shortauthors{Drake et al.}

\begin{document}
\title{Resolving the Nature of the LMC Microlensing Event LMC-5}

\author{A.~J. Drake\altaffilmark{1,2,3}, K.~H. Cook\altaffilmark{3} {\sc and} S.C. Keller\altaffilmark{4}}

\altaffiltext{1}{Dept. of Astrophysical Sciences, Princeton University, Princeton, NJ 08544}
\altaffiltext{2}{Depto. de Astronomia, P. Universidad Catolica, Casilla 104, Santiago 22, Chile}
\altaffiltext{3}{Lawrence Livermore National Laboratory, Livermore, CA 94550}
\altaffiltext{4}{Research School of Astronomy and Astrophysics, ANU, Canberra, ACT 2611, Australia}

\begin{abstract}
 
  We present the results from an analysis of Hubble Space Telescope High
  Resolution Camera data for the Large Magellanic Cloud microlensing event
  MACHO-LMC-5. By determining the parallax and proper motion of this object
  we find that the lens is an M dwarf star at a distance of
  $578^{+65}_{-53}$pc with a proper motion of $21.39 \pm 0.04 $ mas/yr.
  Based on the kinematics and location of this star it more likely to be
  part of the Galactic thick disk than thin disk population. We confirm that
  the microlensing event LMC-5 is a jerk-parallax event.

\end{abstract}

\section{\sc Introduction}

For over a decade astronomers have been observing the Magellanic Clouds in
order to determine the fraction of the dark matter in our Galaxy that may be
in the form of Massive Compact Halo Objects (MACHOs).  The discovery of a
significant number of microlensing events in the first two years of the
MACHO project lead to an uncertain initial estimate that approximately half
of the halo was composed of MACHOs (Alcock et al.~1997). With 3.7 years of
additional data this estimate decreased to $\sim 20\%$ of the halo (Alcock
et al.~2000).  While it appears these objects make up a significant fraction
of the mass in the Galactic halo, little is known about their nature other
than that their most probable masses lie in the range of 0.15 and 0.9$M_\sun$

In order to obtain the most accurate information which can be gained from a
microlensing event, it is useful to accurately determine the flux of the
star that was lensed.  This is made difficult for sources in crowded fields
such as LMC because they are usually blended with neighboring stars.  In
many cases each ``object'' identified in ground-based observations consists
of the blend of a number of stars (Alcock et al.~2000).  The exact location
of the source star is also poorly known because of blending. To determine
the locations of the sources Alcock et al.~(2001a, 2001b) analyzed the MACHO
project images using Difference Image Analysis (Alcock et al.~1999, 2001a).
With these positions they were able to subsequently identify and photometer
the microlensing source stars in observations taken with {\it Hubble Space
  Telescope} (hereafter HST) Wide Field Planetary Camera 2 (WFPC2).

Among the events discovered by the MACHO project toward the Magellanic
clouds was event LMC-5.  This event had a high magnification ($\sim$50) and
was detected in the light curve of Macho object 6.5845.1091 which is located
at $\alpha= 05^{\rm h}\!16^{\rm m}\!41\fs 1$, $\delta =-70\arcdeg\!
29\arcmin\! 18\arcsec$ (J2000).
Gould, Bachall \& Flynn (1997) suggested that the baseline color of this
event was not consistent with an LMC source star and they proposed that the
anomalous color could be attributed to the source being blended with a M
dwarf in the Galactic disk. As the likelihood of finding and M dwarf within
the seeing disk is small they proposed an M dwarf could be the lens in the
foreground. Alcock et al.~(1997) found that the color of the source was in
agreement with the colors and magnitudes of LMC stars, but the event was
indeed blended with a red object.  Most of the LMC microlensing events found
by Alcock et al.~(1997) were blended to some extent.  When the source star
in this microlensing event was identified in HST observations it was
discovered that there was a faint red star nearby.  The probability of
finding an unrelated foreground M dwarf near the microlensing source star is
$\sim 1$ in 10000 (Alcock et al.~2001b).  With this in mind, it was thought
very likely that this object was the lens.

The LMC-5 event shows the clear sign of microlensing parallax. In these
events, the motion of the Earth during the event changes the shape of the
microlensing light curve from the classical Paczy\'nski form (Gould 1992,
Alcock et al.~1995). The presence of parallax enables limits to be placed on
the mass and location of the microlens.  The parallax fit for this event
yielded a lens motion direction that was consistent with the red star having
been the lens.  However, the solution also suggested that the lens was
likely to be a sub-stellar object of 0.036M$_\odot$ ($\leq$ 0.097M$_\odot$
at 3-$\sigma$ significance) at a distance of $\sim 200$pc.  Alcock et
al.~(2001b) derived a separate distance estimate for the lens using the
objects color and spectral type.  First a spectrum of the lens-source
combination was obtained and it was found that the prospective lens was an
M4V or M5V type star.  The V-I color of the object was determined from HST
WFPC2 photometry.  This color was converted into an absolute magnitude using
the $M_{V}$ vs $V-I$ relation of Reid (1991) for M dwarfs. The distance was
then obtained from the observed and absolute magnitudes while the errors in
distance were estimated from the dispersion in M dwarf magnitudes
$V-I\sim3$. The result was that the object was at $650 \pm 190$pc, in stark
contrast to the microlensing parallax solution.

If the red object is indeed the lens, a measurement of the parallax should
confirm the distance inferred from the color of the object in the HST
images.  In addition, a measurement of the magnitude and direction of the
proper motion should agree with the initial estimate which assumed
the red object was the lens, and that the relative separations of 
the two objects represented its proper motion.

To resolve the nature of the candidate lens we undertook a program of
observations with HST's Advanced Camera System High Resolution Camera
(hereafter ACS and HRC, respectively).  In the meantime, a new solution to
the LMC-5 puzzle was proposed by Gould (2004) based on the recent work of
Smith, Mao and Paczy\'nski (2003).  By exploring the phase space of ``vector
microlens parallax'' in a geocentric reference frame, Gould (2004)
discovered a second solution to the microlensing parallax which varied from
the original solution of Alcock et al.~(2001b) by less than 0.1 in fit
$\chi^2$. 

The microlens vector parallax of this second solution differs from that of
the first solution by the so called "jerk parallax", a vector whose
direction lies perpendicular to the direction of the Earth's acceleration
and whose magnitude (for LMC events) is about $(4/3) ({\rm yr/2\pi t_{\rm e}}) \sim 2.4$.  
Events exhibiting these so-called "jerk-parallax degeneracies" are expected
to be rare for microlensing toward the LMC, unless the lens resides in the
Galactic disk.  In the case of the LMC-5 microlensing event, Gould (2004)'s
jerk-parallax solution is in agreement with the lens distance and direction
estimated from the HST photometry.

The solution of Gould (2004) does not rule out the possibility that the
initial solution to the microlensing fit was the correct one, since both are
equally good fits to the lensing light curve.  However, when this solution
is considered in combination with the other evidence from the HST data it is
much more likely that the lens is a sub-stellar object not detected in the
HST data. In this paper we will show with certainty that Gould (2004)'s
solution is the correct one.

\section{\sc Observations}

We obtained images with the HST's HRC in July 2002 and January 2003.  The
observations were taken approximately six months apart to maximize the
parallactic offset of the lens relative to proper motion vector.  Each set
of observations consists of 6 images of the source - lens field, allowing us
to perform robust cosmic ray rejection and to determine very accurate
centroids for each object.  The observations taken in 2002 used the F606W
and F814W filters, while the observations in 2003 were taken in the F606W
filter alone. The duration of each of the exposures was 400 seconds.

\section{\sc Analysis}

The HRC images contain significant distortion in the form of a skew
due to the off-axis location of the ACS. To determine the distortion
corrected location of the stars in the HRC images, we followed the analysis
of Anderson and King (2004). In this process, the standard flat fielded
(flt) HRC images simply were fed into Anderson and King's ``img2xym'' task.
This program finds stars within the images and fits each with an effective
Point Spread Function (ePSF) which is based on an instrumental PSF modified
by the sub-pixel offset of each star's center.
The ePSF varies between observational filters so the correct starting PSF
must be chosen.  The centroid location and flux of each star is determined
in the fitting process. However, because of the large amount of image
distortion the instrumental magnitudes and locations require correction to
an undistorted system where the offset and the changing effective pixel area
are corrected. The ``img2xym'' task also performs these steps to provide
corrected stellar locations and instrumental magnitudes.  The RMS scatter
after the corrections of Anderson and King (2004) is $< 0.01$ pixels, or
about 0.25mas, for the brightest stars in each image.

For each image, we determined the offset between the source star and the
assumed lens. We combined the results for each photometric band separately,
and estimated the uncertainties in these positions based on the scatter in
their locations.  In addition, we combined these locations with the lower
resolution results obtained by Alcock et al.~(2001b) from analysis of HST
WFPC2 observations taken in June 1999. 
It was not possible to estimate the parallax with the prior HST data since
there was only one known location (the WFPC2 point) and one assumed position
at the source star during the microlensing event.


In Table 1 we present a summary of the observations used in this
analysis. We fitted the proper motion and parallax of the object using the
times and locations of the measurements. This fit places the M dwarf at an
offset of ($\Delta X$,$\Delta Y$) $=$ ($2.2\pm9.9$, $-1.6\pm7.8$) mas at the
time of the microlensing event. The reduced $\chi^2$ value of this fit is $<
0.1$. This suggests that the errors in the locations are over estimated, and
the the real uncertainty in this offset if much smaller.  The main
contributor to the uncertainty in the location is the error in the HST WFPC2
location.  However, this result makes it quite certain that the red object
is indeed the lens.  The source star itself is not stationary but moving
with the proper motion of the LMC which is  
($\mu_{\alpha {\rm LMC}}$, $\mu_{\delta {\rm LMC}}$) = (1.68, 0.34) mas/yr
(van der Marel et al.~2002).  However, in the case of microlensing events we
are only interested in the motion of the lens relative to the source.

With the lens identified, an additional constraint for determining the
proper motion and trigonometric parallax of the lens was derived from the
fact that we know that the source star and the lens must be collinear in our
line-of-sight at the time of the microlensing event peak amplification. We
fitted the locations again to determine the proper motion and parallax of
the lens with the inclusion of this additional point.  We find the proper
motion of the lens relative to the source to be 
($\mu_{\alpha {\rm SL}}$, $\mu_{\delta {\rm SL}}$) = ($17.56 \pm 0.04$, $-12.22\pm0.02$ ) mas/yr.
The position angle of the proper motion vector is $\theta = 124.8\arcdeg$.
This in exellent agreement with the direction of Gould (2004)'s solution
of $123.9\arcdeg$.  We note that the direction of proper motion in ecliptic
coordinates was incorrectly given by Alcock et al.~(2001b) as $\theta_{sky} = -91.6\arcdeg$, 
rather than $\theta_{sky} =-105.7\arcdeg$. It appears that $\Delta\lambda$ was used
to determine the direction of motion instead of $\rm \Delta\lambda cos(\beta)$.

We find the parallax of the lens to be $\pi_{\rm L} = 1.73\pm0.18$ mas.
Therefore, the lens lies at a distance of $578^{+65}_{-53}$ pc.  This result
is in agreement with the previous photometric estimate of Alcock et
al.~(2001b) ($650$ pc).  The fact that we have been able to measure a $\sim
2$ mas parallax to $\sim$10\% uncertainty is a good demonstration of the
astrometric accuracy that can be achieved with the HST HRC instrument.


\begin{figure}[ht]
\epsscale{1.1}
\plotone{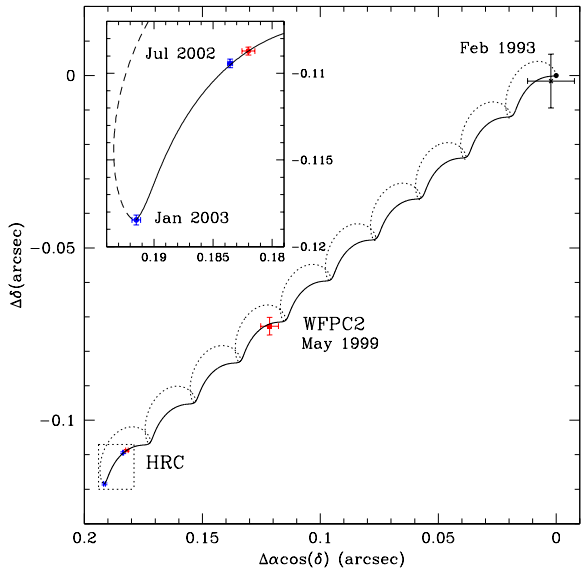}  
\figcaption{\label{lmc5}   
  Motion of the microlens relative to the position MACHO source star
  6.5845.1091 at the time of the microlensing peak magnification. The insert
  presents an expanded view of dotted region where locations were determined
  with the HRC instrument.
The dot at (0,0) shows the location of the source star and the large errors
bars near (0,0) shows the position of the object at the time of the
microlensing event determined from the initial parallax and proper motion
fit. See the text for further details.}
\end{figure}

In Figure \ref{lmc5} we present the fit to the motion of the LMC-5 lens
corrected for the source star motion.  The solid line shows the fit to the
HST data including the source star location as a point, while the dashed
line shows the proper motion and parallax that is expected with the lens
distance of 200pc as determined from the original microlensing parallax fit.
The fit shown in this figure is slightly more constrained than it appears
since the times of the measurements are an important part of the fitting
process. The F606W and F814W HRC points taken in July 2002 should lie very
close to each other. However, as the figure shows they are significanty
offset. We have checked for systematic errors in the transformations of
Anderson and King (2004) by matching a large number of stars between the
F606W and F814W frames. The coordinates of these stars matched within the
centroid uncertainties while the lens appeared to be slightly offset between
bands.  However, the current level of uncertainty is too large to tell
whether the offset is real or due to an unquantified localized distortion.

If we assume the lens undergoes average Galactic foreground reddening for
the LMC of E(B-V) = 0.06 (Oestreicher, Gochermann, Schmidt-Kaler 1995), we
find $M_{V}= 13.68$, consistent with an M5 dwarf star. 
Our results are in agreement with the spectra and $M_{V}$ presented by
Alcock et al.~(2001b).  It is very difficult to observationally
rule out the possibility that a 0.036M$_\odot$ object at 200pc as the lens
since the object could be fainter than 30th magnitude in V band (Baraffe et al.~1998).
This mass lies below the limits where models are accurate. However, K band 
observations may be more promising.
 The fact that we know that foreground M dwarfs are extremely
rare in fields toward the LMC (Alcock et al.~2001b) suggests this object is
very likely the lens. From our initial parallax and proper motion fit we
known that the object was within a few milli-arcseconds of the source at the
time of the lensing event.  Our results agree with the new jerk-parallax
solution discovered by Gould (2004).  Therefore, there is almost no doubt
that the M dwarf we have observed is the indeed the lens.

For the lens we find the space velocity components (U,V,W) to be
(43.2,-55.7,29.0) $\rm km.s^{-1}$ corrected to the Local Standard of Rest.
These values are much higher than those presented by Alcock et al.~(2001b)
as they assumed a distance of 200pc and are quite high for thin disk M
dwarfs. However, the displacement from the Galactic plane ($\sim$300pc) and
the velocity are consistent with both thin and thick disk stars. We have
simulated the stellar population of disk and halo stars toward the LMC field
following the method of Vallenari, Bertelli, and Schmidtobreick (2000).
Using a common range of disk parameters (scale height, scale length, etc.).
We find that the lens is slightly more likely to be a thick disk star
($\sim 50\pm30\%$) than a thin disk one.  Clearly the likelihood is strongly
dependent on adopted parameters for the Galactic components.  The kinematics
are also in good agreement with the thick disk kinematics derived by Chiba
and Beers (2000).

\section{Conclusions}

We have analyzed HRC data for LMC microlensing event LMC-5 and we find that
the lens in this microlensing event is an M dwarf star.  Based on our
analysis we can confirm that the jerk-parallax solution to the microlensing
light curve discovered by Gould (2004) is correct.  The kinematics of this
star suggest that it is most likely a part of the Galactic thick disk
population rather than part of the dark halo. This is the first time that
any microlens has been identified with such certainty.  However, this
discovery does not affect the current estimates of the mass fraction of the
Galactic dark halo in the form of MACHOs, since some microlensing events due
to foreground disk stars are expected in all LMC microlensing models.

We would like to thank an anonymous referee for his many helpful suggestions.
We would also like to thank Jay Anderson who generously made his
results and analysis programs available to us prior to their release.
Support for this publication was provided by NASA through proposal numbers
GO-9394 and from the Space Telescope Science Institute, which is operated by
the Association of Universities for Research in Astronomy, under NASA
contract NAS5-26555.
This work was performed under the auspices of the U.S.~Department of Energy
National Nuclear Security Administration by the University of California,
Lawrence Livermore National Laboratory under contract W-7405-Eng-48.  The
work done by A.~J.~D. is supported by Chilean FONDECYT grant 1030955.


\begin{deluxetable}{llll}
\tablecaption{Astrometric data for LMC-5.\label{tab1}}
\tablewidth{0pt}
\tablehead{\colhead{Observations} &  \colhead{$\Delta \alpha$} & \colhead{$\Delta \delta$} & \colhead{Time}\\
\colhead{} &  \colhead{$\arcsec$} & \colhead{$\arcsec$} & \colhead{days}}
\startdata
WFPC2 (F555W \& F814W)  &  $0.1110\pm0.0038$ & $-0.0748\pm0.0026$ & 2288.16\nl
ACS HRC (F606W) &  $0.16772\pm0.00021$ & $-0.11263\pm0.00024$ & 3442.55\nl
ACS HRC (F814W) &  $0.16618\pm0.00055$ & $-0.11192\pm0.00023$ & 3442.64\nl
ACS HRC (F606W) &  $0.17484\pm0.00037$ & $-0.12183\pm0.00028$ & 3621.52
\enddata
\tablecomments{
Col. (1), instrument and filters used in observations.
Cols. (2) \& (3), relative offsets between source and lens in 
right ascension (great-circle) and declination, respectively.
Col. (4), observation time relative to peak time of lensing event 
(JD=2449023.9).
}
\end{deluxetable}

\end{document}